\documentstyle[12pt]{article}

\begin{document}
\title{Lie symmetries for two-dimensional charged particle motion}
\author{F.~Haas and J.~Goedert \\
Centro de Ci\^encias Exatas e Tecnol\'ogicas, UNISINOS\\
Av. Unisinos, 950\\
93022--000 S\~ao Leopoldo, RS - Brazil}
\date{\strut}
\maketitle

\begin{abstract}
We find the Lie point symmetries for non--relativistic
two-di\-men\-sio\-nal charged particle motion. These symmetries
comprise a qua\-si--in\-va\-rian\-ce transformation, a
time--dependent rotation, a time--dependent spatial translation
and a dilatation. The associated electromagnetic fields satisfy a
system of first--order linear partial differential equations. This
system is solved exactly, yielding four classes of electromagnetic
fields compatible with Lie point symmetries.
\end{abstract}

\section{Introduction}

Lorentz equations for non--relativistic charged particle motion
constitute a very basic dynamical system whose symmetry structure
deserves a detailed investigation. In a previous paper
\cite{Haas}, we have studied the Noether point symmetries for
two--dimensional non--relativistic charged particle motion. Here
we make a systematic search of the Lie point symmetries associated
with two-- dimensional non--relativistic charged particle motion
under general electromagnetic fields. The reasons for this are
two--fold. First, as is well known \cite{Sarlet}, the Lie point
symmetry group is more general and contains the Noether point
symmetry group of any given problem. In fact, the Lie point
symmetries extends the class of Noether point symmetries of planar
non--relativistic charged particle motion by the incorporation of
an extra scale symmetry. Second, even if the Lie point symmetry
approach does not give first integrals as directly as Noether's
theorem, it does opens the possibility for reducing the number of
relevant variables in the system. Reduction of variables is of
importance in reducing the amount of computational time in
numerical simulations. For example, in the numerical treatment of
the Vlasov--Maxwell system in colisionless plasma physics
\cite{Lewis, Schrauner}, it is highly desirable to know the most
general electromagnetic field configurations having Lie point
symmetry.

Herein, we consider planar non--relativistic charged particle
motion under initially general electromagnetic field. The
corresponding Lorentz equations read
\begin{eqnarray}
\label{eq1} \ddot x &=& E_{1}(x,y,t) + \dot{y}\,B(x,y,t) \,,\\
\label{eq2} \ddot y &=& E_{2}(x,y,t) - \dot{x}\,B(x,y,t) \,,
\end{eqnarray}
where ${\bf E} = (E_{1}(x,y,t), E_{2}(x,y,t), 0)$ is the planar
electric field and ${\bf B} = (0, 0, B(x,y,t))$ is the
perpendicular magnetic field. Unlike the usual approach to Lie
symmetries for charged particle motion \cite{Ritter}, we do not
start with a prescribed electromagnetic field. Rather, we find the
conditions on ${\bf E}$ and ${\bf B}$ so that the system
(\ref{eq1}--\ref{eq2}) do admit a Lie point symmetry. With this
strategy in mind we do not stick to any particular electromagnetic field
but search for the most general form that may present the symmetry.
Once the general forms are known they can eventually be
specified in more detail to fit some particular application. In
fact, electromagnetic fields written in terms of arbitrary
functions are of fundamental importance for treating the
Vlasov--Maxwell system in colisionless plasma physics \cite{Lewis,
Schrauner}. Therefore, we do not consider, in the continuation, any
choice implying a too much particular field configuration, like
that homogeneous in space.  This strategy excludes some of the Lie
point symmetries. Nevertheless, the approach gives the most
general electromagnetic fields containing a number of arbitrary
functions, that can be fixed later to fit particular applications
like those of plasma physics.

As in the Noether point symmetry case \cite{Haas}, the Lie point
symmetry approach also produces a system of linear, first order
partial differential equations to be satisfied by the
electromagnetic fields. We solve this system using a
strategy similar to that used in reference \cite{Haas}. That is, we find
the canonical group coordinates for the various Lie symmetries and
express the resulting system of partial differential equations
in these coordinates. As a
consequence, the system is transformed in a set of ordinary
differential equations that can be solved easily. Our procedure
shows how the problem of finding classes of electromagnetic
fields compatible with Lie point symmetries is equivalent to the
problem of finding canonical group variables for these same
symmetries.

The paper is organized as follows. In section 2, we obtain the
most general form of the Lie point symmetries associated with
planar, non--relativistic charged particle motion. In the same
section, we obtain the system of equations satisfied by the
corresponding electromagnetic field. Section 3 is dedicated to the
calculation of the canonical group variables for the
Lie point symmetries. We find four classes of canonical
coordinates. In section 4, the basic system of partial
differential equations satisfied by the electromagnetic field is
solved for each of the four classes of canonical group
variables. Section 5 is devoted to the conclusions.

\section{Lie point symmetries}

Let us consider infinitesimal point transformations,
\begin{eqnarray}
\label{eq3} \bar{x} &=& x + \varepsilon\eta_{1}(x,y,t) \,,\\
\label{eq4} \bar{y} &=& y + \varepsilon\eta_{2}(x,y,t) \,,\\
\label{eq5} \bar{t} &=& t + \varepsilon\tau(x,y,t) \,,
\end{eqnarray}
where $\varepsilon$ is an infinitesimal parameter. For future
convenience we denote the generator of the group of symmetries
associated to (\ref{eq4}--\ref{eq6}) by
\begin{equation}
\label{eq6} G = \tau\frac{\partial}{\partial t} +
\eta_{1}\frac{\partial}{\partial x} +
\eta_{2}\frac{\partial}{\partial y} \,.
\end{equation}
The generator $G$  appears frequently in what follows and is
useful in the definition of canonical group coordinates, which
plays a central role in the systematic determination of the
electromagnetic fields associated the symmetries.

The condition for Lie symmetry \cite{Bluman,
Leach} reads, in this case
\begin{equation}
\label{eq7} G^{[2]}\left({\bf N}\right)_{{\bf N}=0} = 0 \,,
\end{equation}
where ${\bf N} = (N_{1},N_{2})$,
\begin{eqnarray}
N_1 &=& \ddot{x} - E_{1}(x,y,t) - \dot{y}\,B(x,y,t) \,,\\ N_2 &=&
\ddot{y} - E_{2}(x,y,t) + \dot{x}\,B(x,y,t) \,,\end{eqnarray}
$G^{[2]}$ being the generator of the twice extend group
(for its complete expression see, for instance,
references \cite{Bluman, Leach}). The calculation of Lie point symmetries
is a fairly known procedure and we only outline its main steps
here. By inserting the equations of motion into the Lie symmetry
criteria (\ref{eq7}) we obtain a polynomial equation in the velocity
components. For this polynomial form, condition (\ref{eq7}) implies that the
coefficients of all monomials of the form ${\dot x}^{m}{\dot y}^n$
must be identically zero. This yields a system of
partial differential equations to be satisfied by $\tau, \eta_1$
and $\eta_2$. For instance, the terms cubic in velocity give a
system with general solution
\begin{equation}
\label{eq8} \tau = \rho^{2}(t) + g_{1}(t)\,x + g_{2}(t)\,y \,,
\end{equation}
where $g_1$,$\dots$,$g_6$ are arbitrary functions of the indicated
arguments. Equation (\ref{eq8}) wil be taken into account in the
continuation.

The terms quadratic in the velocity yield
\begin{eqnarray}
\label{eq9} \eta_{1xx} - 2\dot{g}_1 + g_{2}\,B &=& 0 \,, \qquad
\eta_{1xy} - \dot{g}_2 - g_{1}\,B = 0 \,, \\ \label{eq10}
\eta_{1yy} - g_{2}\,B &=& 0 \,, \qquad \eta_{2xx} + g_{1}\,B = 0 \,, \\
\label{eq11}
\eta_{2xy}- \dot{g}_1 + g_{2}\,B &=& 0 \,, \qquad \eta_{2yy} - 2\dot{g}_2 -
g_{1}\,B = 0 \,,
\end{eqnarray}
where we have used subscripts to denote partial derivatives.
Direct inspection shows that the choice
\begin{equation}
\label{eq12}
g_1 = g_2 = 0
\end{equation}
keeps $B$ arbitrary and
implies that $\eta_1$ and $\eta_2$ are linear functions of position,
\begin{eqnarray}
\label{eq13} \eta_1 &=& g_{3}(t)\,x + g_{4}(t)\,y + a_{1}(t) \,,
\\ \label{eq14} \eta_2 &=& g_{5}(t)\,x + g_{6}(t)\,y + a_{2}(t)
\,,
\end{eqnarray}
where $g_3$, $g_4$, $g_5$, $g_6$, $a_1$ and $a_2$ are functions of
time only.  It is now worth to stress that the choice (\ref{eq12})
implies no restriction on the magnetic field, which remains
completely arbitrary. Moreover, a detailed calculation involving
equations (\ref{eq9}--\ref{eq11}) shows that the only way of
keeping spatial dependence in the magnetic field is to set $g_1 =
g_2 = 0$.  In fact, we are interested in classes of magnetic fields
more general than simply those homogeneous in space. Hence, we
adopt (\ref{eq12}), with the corresponding solution
(\ref{eq13}--\ref{eq14}) for $\eta_1$ and $\eta_2$.
The important point here is that, up to this stage, we have
preserved the complete
arbitrariness of the magnetic field.

The terms linear in velocity yield
\begin{eqnarray}
\label{eq18} (g_4 + g_{5})\,B &=& - 2(\rho\ddot\rho + \dot\rho^2)
+ 2\dot{g}_3 \,,\\ \label{eq19} (g_4 + g_{5})\,B &=&
2(\rho\ddot\rho + \dot\rho^2) - 2\dot{g}_6 \,,\\ \label{eq20} G\,B
&=& (g_3 - g_6 - 2\rho\dot\rho)\,B + 2\dot{g}_4 \,,\\ \label{eq21}
G\,B &=& (g_6 - g_3 - 2\rho\dot\rho)\,B - 2\dot{g}_{5} \,,
\end{eqnarray}
where $G$ is the generator defined in (\ref{eq6}).

An examination of equations (\ref{eq18}--\ref{eq19}) shows that
the only way of preserving space dependence in the magnetic
field is to set
\begin{equation}
\label{eq22} g_4 = - g_{5} = - \Omega(t) \,,
\end{equation}
where $\Omega$ is a function of time only. Moreover, this
implies, from (\ref{eq18}--\ref{eq19}),
\begin{equation}
\label{eq23} g_3 = \rho\dot\rho + k_1 \,, \qquad g_6 =
\rho\dot\rho + k_2 \,,
\end{equation}
with, $k_1$ and $k_2$ numerical constants. Equation
(\ref{eq23}) and compatibility between equations (\ref{eq20}--\ref{eq21})
give
\begin{equation}
\label{eq24} (k_1 - k_{2})\,B = 0 \,.
\end{equation}
For $B$ not identically zero, the conclusion is
\begin{equation}
\label{eq25} k_1 = k_2 = k \,,
\end{equation}
where $k$ is a numerical constant. Equations
(\ref{eq20}--\ref{eq21}) now furnishes
\begin{equation}
\label{eq26} G\,B = - 2\rho\dot\rho\,B - 2\dot\Omega \,,
\end{equation}
where the generator of Lie symmetries may be split into
four components
\begin{equation}
\label{eq27} G = G_Q + G_R + G_T + G_S\,.
\end{equation}
In equation (\ref{eq27}),
\begin{equation}
\label{eq28} G_Q = \rho^{2}(t)\frac{\partial}{\partial t} +
\rho\dot\rho\,(x\frac{\partial}{\partial x} +
y\frac{\partial}{\partial y}) \quad ,
\end{equation}
is the generator of quasi--invariance transformations \cite{Munier},
\begin{equation}
\label{eq29} G_R =  \Omega(t)\,(x\frac{\partial}{\partial y} -
y\frac{\partial}{\partial x})
\end{equation}
generates time--dependent rotations,
\begin{equation}
\label{eq30} G_T = a_{1}(t)\frac{\partial}{\partial x} +
      a_{2}(t)\frac{\partial}{\partial y}
\end{equation}
generates time--dependent spatial translations and
\begin{equation}
\label{eq31} G_S = k\left(x\,\frac{\partial}{\partial\,x} +
y\,\frac{\partial}{\partial\,y}\right)
\end{equation}
is the generator of dilatations (or, more precisely, of
contractions if $k < 0$). Comparison with the generator of Noether
point symmetries for two--dimensional non--relativistic
charged--particle motion \cite{Haas} shows that the Lie point symmetry
generator has an additional term, depending on the parameter $k$.
We also observe that this form is essentially new, and
cannot be expressed in terms of the generators $G_Q$, $G_R$ and
$G_T$.

The terms independent of velocity in the Lie invariance condition
have also to be taken into account. They produce the equations
for the electric field,
\begin{eqnarray}
G\,E_1 = (- 3\rho\dot\rho &+& k)\,E_1 - \Omega\,E_2 -
\left((\rho\ddot\rho +
\dot\rho^2)y + \dot\Omega\,x + \dot{a}_2\right)\,B \nonumber + \\
\label{eq32} &+& (\rho{\buildrel \cdots\over\rho} +
3\dot\rho\ddot\rho)\,x - \ddot\Omega\,y + \ddot{a}_1 \,, \\ G\,E_2
= (- 3\rho\dot\rho &+& k)\,E_2 + \Omega\,E_1 +
\left((\rho\ddot\rho + \dot\rho^2)x - \dot\Omega\,y +
\dot{a}_{1}\right)\,B + \nonumber
\\ \label{eq33} &+& (\rho{\buildrel \cdots\over\rho} +
3\dot\rho\ddot\rho)\,y + \ddot\Omega\,x + \ddot{a}_2 \,.
\end{eqnarray}

Let us summarize the results obtained so far. Implicitly, in our
treatment, we excluded the excessively restricted class of
spatially homogeneous magnetic fields depending on time only. This
approach yields the system equations (\ref{eq26}),
(\ref{eq32}--\ref{eq33}), which must be satisfied by the electromagnetic fields
associated to Lie point symmetries of the planar charged particle
motion. The Lie symmetry generator in equations (\ref{eq26}),
(\ref{eq32}--\ref{eq33}), which constitute a system of linear,
first order, coupled partial differential equations for $E_1$,
$E_2$ and $B$, is given by (\ref{eq27}). In comparison with the
treatment of two--dimensional non--relativistic charged particle
motion with Noether point symmetries, we find that Lie point
symmetries have an extra component associated to scale
transformation. This extra contribution modifies both the generator
of symmetries and the equations satisfied by the electromagnetic
fields \cite{Haas}.

In the remaining of this paper, we are essentially concerned with
finding all the solutions of the system of partial differential
equations (\ref{eq26}), (\ref{eq32}--\ref{eq33}). These solutions
yield the most general electromagnetic field under which the
planar motion of charged particles present Lie point symmetry. An
useful remark is that $B$ satisfies an equation decoupled from the
equations for $E_1$ and $E_2$, whereas the equations for the
electric field do depend on $B$. Thus, we must first solve
(\ref{eq26}) for $B$ and only afterwards treat
(\ref{eq32}--\ref{eq33}) for the electric field.

Finally, if
these solutions should constitute true electromagnetic fields,  the
additional requirement of Faraday's law
\begin{equation}
\label{eq34} E_{2x} - E_{1y} + B_t = 0 \,,
\end{equation}
must be imposed, to comply with Maxwell's equations.

To treat the system (\ref{eq26}), (\ref{eq32}--\ref{eq33}) and to
find its complete solution, we shall use canonical group coordinates.
These variables are introduced in the section that follows.

\section{Canonical group coordinates}

Canonical group coordinates \cite{Bluman, Leach} are defined by imposing
that the symmetry transformation behaves merely like  time
translation. Denoting new coordinates by
$(\bar{x},\bar{y},\bar{t})$, this means that, in
canonical group coordinates,
\begin{equation}
\label{eq35} G = \frac{\partial}{\partial\bar{t}} \,,
\end{equation}
where $\bar{t}$ is the new time parameter. This implies that
canonical group coordinates satisfy the equations
\begin{equation}
\label{eq36} G\,\bar{x} = 0 \,, \qquad G\,\bar{y} = 0 \,,
\qquad G\,\bar{t} = 1 \,.
\end{equation}
This set of uncoupled linear partial differential equations, can be
solved, for the generator (\ref{eq27}), in closed
form by the method of characteristics. We find four classes of
solutions, listed below.

\subsection{The case $\rho \neq 0$}

When $\rho \neq 0$, it is convenient to write
\begin{eqnarray}
\label{eq37} a_1 &=& \rho^{2}\dot\alpha_{1} - (\rho\dot\rho + k)
\alpha_{1} \,,\\ \label{eq38} a_2 &=& \rho^{2}\dot\alpha_{2} -
(\rho\dot\rho + k)\alpha_{2}
\end{eqnarray}
for suitable functions $\alpha_{1}(t)$ and $\alpha_{2}(t)$, which are
therefore defined in terms of $a_1$ and $a_2$.

In terms of (\ref{eq37}--\ref{eq38}), we have the
following canonical group coordinates,
\begin{eqnarray}
\label{eq39} \bar{t} &=& \int^{t}d\mu/\rho^{2}(\mu) \,,\\
\label{eq40} \bar{x} &=&
\frac{e^{-k\bar{t}}}{\rho}\left((x-\alpha_{1})\cos\,T +
(y-\alpha_{2})\sin\,T\right) + \delta_{1}  \,,\\ \label{eq41}
\bar{y} &=&
\frac{e^{-k\bar{t}}}{\rho}\left(-
(x-\alpha_{1})\sin\,T + (y-\alpha_{2})\cos\,T \right) + \delta_{2}
\,,
\end{eqnarray}
where new functions $T = T(t)$, $\delta_1 = \delta_{1}(t)$ and
$\delta_2 = \delta_{2}(t)$ were defined according to
\begin{eqnarray}
\label{eq42} T(t) &=& \int^{t}d\mu\,\Omega(\mu)/\rho^{2}(\mu)
\,,\\ \label{eq43} \delta_{1}(t) &=& - \int^{t}d\mu
\frac{\Omega(\mu)}{\rho^{3}(\mu)}e^{-k\bar{t}(\lambda)}\left(
\alpha_{1}(\mu)\sin\,T(\mu) - \alpha_{2}(\mu)\cos\,T(\mu)\right)
\,,\\ \label{eq44} \delta_{2}(t) &=& - \int^{t}d\mu
\frac{\Omega(\mu)}{\rho^{3}(\mu)}e^{-k\bar{t}(\lambda)}\left(\alpha_{1}(\mu)
\cos\,T(\mu) + \alpha_{2}(\mu)\sin\,T(\mu)\right) \,.
\end{eqnarray}
Notice that the requirement $\rho \neq 0$ is essential, for
otherwise the canonical group variables (\ref{eq39}--\ref{eq41})
would not be well defined.

For $k = \Omega = \alpha_1 = \alpha_2 = 0$, the transformation
(\ref{eq39}--\ref{eq41}) is known as the quasi-invariance
transformation \cite{Munier}. In the general case, however, the
transformations includes also
dilatation, time--dependent rotation and time dependent
translation. In comparison with the Noether point symmetry
approach, the present set of canonical group variables are in
direct correspondence with case 3.1 of reference \cite{Haas}. In
fact, when $k=0,$ formulae (\ref{eq39}--\ref{eq41}) become formulae (42--44)
of \cite{Haas}.

\subsection{The case $\rho = k = 0$ and $\Omega \neq 0$}

In this case, we have Noether point symmetry \cite{Haas}. The
canonical group variables are
\begin{eqnarray}
\label{eqq44} \bar{t} &=& \frac{1}{\Omega}\tan^{-1}
\left(\frac{y - \beta_{2}}{x - \beta_{1}}\right) \,, \\
\label{eqq42} \bar{x} &=& \left((x - \beta_{1})^2 + (y -
\beta_{2})^{2}\right)^{1/2} \,,\\ \label{eqq43} \bar{y} &=& t
\,,
\end{eqnarray}
with
\begin{equation}
\label{eqqq44} \beta_1 = \beta_{1}(t) = - a_{2}/\Omega \quad ,
\quad \beta_2 = \beta_{2}(t) = a_{1}/\Omega \quad .
\end{equation}
The variables $\bar{x}$ and $\bar{t}$ are translated polar
coordinates, the new time parameter playing the role of an azimuthal
angle and $\bar{x}$ the role of a radial coordinate.

\subsection{The case $\rho = k = \Omega = 0$ and $a_{2} \neq 0$}

Again, we have Noether point symmetry with canonical group
variables
\begin{eqnarray}
\label{eqq47} \bar{t} &=& y/a_2 \,,\\
\label{eqq45} \bar{x} &=& x - a_{1}y/a_{2} \,,\\ \label{eqq46}
\bar{y} &=& t \,.
\end{eqnarray}
We finally mention that the case $\rho = 0$, $\Omega = 0$ and
$a_{1} \neq 0$ is strictly analogous to this last case and deserves
no special consideration.

\subsection{The case $\rho = 0$, $k \neq 0$.}

In this case the canonical coordinates are
\begin{eqnarray}
\label{eq45}
\bar{t} &=& \frac{1}{2k}\log\left((x - \gamma_1)^2 + (y -
\gamma_2)^2\right) \,,\\
\label{eq46}
\bar{x} &=& \arctan\left(\frac{y - \gamma_2}{x - \gamma_1}\right) -
\Omega\bar{t} \,,\\
\label{eq47}
\bar{y} &=& t \,,
\end{eqnarray}
where
\begin{equation}
\label{eq48}
\gamma_1 = - \frac{k\,a_1 + \Omega\,a_2}{k^2 + \Omega^2} \,,\qquad \gamma_2 =
\frac{\Omega\,a_1 - k\,a_2}{k^2 + \Omega^2} \,.
\end{equation}

The symmetry transformation represents a dilatation, plus a
ti\-me--de\-pen\-dent rotation and a time--dependent translation.

\section{Electromagnetic fields}

We now tackle equations (\ref{eq26}), (\ref{eq32}--\ref{eq33})
for the electromagnetic fields in each of the four possible
symmetry transformations expressed in canonical group
variables.

\subsection{The case $\rho \neq 0$}

Equation (\ref{eq26}), which involves only the magnetic field
acquires, in canonical group coordinates, the form
\begin{equation}
\label{eq49}
B_{\bar{t}} = - \frac{2\rho'}{\rho}B - \frac{2\Omega'}{\rho^2} \,,
\end{equation}
where prime denotes total differentiation with respect to $\bar{t}$.
The general solution for (\ref{eq49}) is
\begin{equation}
\label{eq50}
B = - \frac{2\Omega}{\rho^2} + \frac{1}{\rho^2}\bar{B}(\bar{x},\bar{y}) \,,
\end{equation}
where $\bar{B}(\bar{x},\bar{y})$ is an arbitrary function of the
indicated arguments. Notice that the resulting magnetic field is
not necessarily homogeneous, since it can depend on the spatial
coordinates through $\bar{x}$ and $\bar{y}$. This is a significant
improvement on earlier results \cite{Bouquet}.
Formally, $B$ is identical to the magnetic field of case 4.1 in
reference \cite{Haas} on Noether point symmetries. Notice,
however, the different form of canonical group variables.

To find the corresponding electric field, we must solve the system
(\ref{eq32}--\ref{eq33}), taking the solution (\ref{eq50}) into
account. In this case, it is useful to introduce the quantities
$\Sigma_1$ and $\Sigma_2$ defined by
\begin{eqnarray}
\label{eq51}
\Sigma_1 &=& \rho^{3}e^{- k\bar{t}}(E_{1}\cos\,T + E_{2}\sin\,T) \,,\\
\label{eq52}
\Sigma_2 &=& \rho^{3}e^{- k\bar{t}}(- E_{1}\sin\,T + E_{2}\cos\,T) \,,
\end{eqnarray}
representing a rotation plus a rescaling of the electric field
which, in this case, can be viewed as of a circularly polarized
wave with time-dependent amplitude.  In the new variables, the
system (\ref{eq32}--\ref{eq33}) decouples and can be cast into the
form
\begin{equation} \label{eq53}
\frac{\partial\Sigma_1}{\partial\bar{t}} =
\frac{\partial\psi_1}{\partial\bar{t}} \quad , \quad
\frac{\partial\Sigma_2}{\partial\bar{t}} =
\frac{\partial\psi_2}{\partial\bar{t}} \quad , \end{equation}
where
\begin{eqnarray} \psi_1 &=& \left(-
\frac{\rho'}{\rho}(\bar{y} - \delta_{2}) + \delta_{2}' + k\delta_2
- \Omega(\bar{x} - \delta_{1}) + \frac{e^{- k\bar{t}}}{\rho}
(\alpha_{1}'\sin\,T - \alpha_{2}'\cos\,T)\right)
\bar{B}(\bar{x},\bar{y}) + \nonumber \\
&+&
\left(\frac{\rho''}{\rho} - 2\frac{{\rho'}^2}{\rho^2} +
\Omega^{2}\right)(\bar{x} - \delta_{1}) - \left(\Omega' -
2\frac{\rho'}{\rho}\Omega\right)(\bar{y} - \delta_{2}) + \nonumber
\\ &+& \frac{e^{- k\bar{t}}}{\rho}\left(\Omega'\alpha_1 -
\Omega(\alpha_{1}' + \frac{\rho'}{\rho}\alpha_1) + \alpha_{2}'' -
2\frac{\rho'}{\rho}\alpha_{2}'
 + \Omega^{2}\alpha_{2}\right)\sin\,T + \\
&+& \frac{e^{- k\bar{t}}}{\rho}\left(- \Omega'\alpha_2 + \Omega(\alpha_{2}' +
\frac{\rho'}{\rho}\alpha_2) + \alpha_{1}'' - 2\frac{\rho'}{\rho}\alpha_{1}'
 + \Omega^{2}\alpha_{1}\right)\cos\,T \nonumber \\
&-& k\,(\delta_{1}' + k\delta_1)
 \,, \nonumber \\
&\strut& \nonumber \\
\psi_2 &=& \left(+
\frac{\rho'}{\rho}(\bar{x} - \delta_{1}) - \delta_{1}' - k\delta_1 -
\Omega(\bar{y} - \delta_{2}) + \frac{e^{- k\bar{t}}}{\rho}
(\alpha_{1}'\cos\,T
+ \alpha_{2}'\sin\,T)\right) \bar{B}(\bar{x},\bar{y}) + \nonumber
\\ &+& \left(\frac{\rho''}{\rho} - 2\frac{{\rho'}^2}{\rho^2} +
\Omega^{2}\right)(\bar{y} - \delta_{2}) + \left(\Omega' -
2\frac{\rho'}{\rho}\Omega\right)(\bar{x} - \delta_{1})  \nonumber
\\ &-& \frac{e^{- k\bar{t}}}{\rho}\left(- \Omega'\alpha_2 +
\Omega(\alpha_{2}'
+ \frac{\rho'}{\rho}\alpha_2) + \alpha_{1}'' -
2\frac{\rho'}{\rho}\alpha_{1}'
 + \Omega^{2}\alpha_{1}\right)\sin\,T + \\
&+& \frac{e^{- k\bar{t}}}{\rho}\left(+ \Omega'\alpha_1 - \Omega(\alpha_{1}' +
\frac{\rho'}{\rho}\alpha_{1}) + \alpha_{2}'' - 2\frac{\rho'}{\rho}\alpha_{2}'
 + \Omega^{2}\alpha_{2}\right)\cos\,T \nonumber \\
&-& k\,(\delta_{2}' +  k\delta_2) \,.
\nonumber
\end{eqnarray}

The general solution for (\ref{eq53}) is
\begin{equation}
\label{eq54}
\Sigma_1 = \psi_1 + \bar{E}_{1}(\bar{x},\bar{y}) \quad , \quad
\Sigma_2 = \psi_2 + \bar{E}_{2}(\bar{x},\bar{y}) \,,
\end{equation}
where, as indicated, $\bar{E}_1$ and $\bar{E}_2$ have no
dependence on $\bar{t}$.

We are interested in the electric field, in the original variables. To
obtain the field in this coordinates
we use the inverse of the transformation (\ref{eq51}--\ref{eq52}),
\begin{eqnarray}
\label{eq55}
E_1 &=& \frac{e^{k\bar{t}}}{\rho^3}(\Sigma_{1}\cos\,T - \Sigma_{2}\sin\,T) \,,\\
\label{eq56}
E_2 &=& \frac{e^{k\bar{t}}}{\rho^3}(\Sigma_{1}\sin\,T + \Sigma_{2}\cos\,T) \,.
\end{eqnarray}
Substituting equations (\ref{eq55}--\ref{eq56}) into (\ref{eq54}) and
transforming back to the original
variables $(x,y,t)$, yields the electric field components
\begin{eqnarray}
E_1 &=& \ddot\alpha_1 + \frac{\ddot\rho}{\rho}(x - \alpha_{1}) +
\frac{\Omega^{2}x}{\rho^4} - (\rho\dot\Omega -
2\dot\rho\Omega)\frac{y}{\rho^3} +
\frac{\Omega}{\rho^3}(\rho\dot\alpha_2 - \dot\rho\alpha_{2}) +
\nonumber \\
&+& \frac{k^{2}e^{k\bar{t}}}{\rho^3}(\delta_{2}\sin\,T - \delta_{1}\cos\,T)
 - \frac{k\Omega\alpha_2}{\rho^4} + \nonumber \\
\label{eq57} &+&
\frac{e^{k\bar{t}}}{\rho^3}\left(\bar{E}_{1}(\bar{x},\bar{y})\cos\,T
- \bar{E}_{2}(\bar{x},\bar{y})\sin\,T\right) \\ &-&
\frac{1}{\rho^4}\left(\rho\dot\rho(y - \alpha_2) +
\rho^{2}\dot\alpha_2 + \Omega\,x -
k\rho\,e^{k\bar{t}}(\delta_{2}\cos\,T +
\delta_{1}\sin\,T)\right)\bar{B}(\bar{x},\bar{y}) \,, \nonumber \\
&\strut& \nonumber \\ E_2 &=& \ddot\alpha_2 +
\frac{\ddot\rho}{\rho}(y - \alpha_{2}) +
\frac{\Omega^{2}y}{\rho^4} + (\rho\dot\Omega -
2\dot\rho\Omega)\frac{x}{\rho^3} -
\frac{\Omega}{\rho^3}(\rho\dot\alpha_1 - \dot\rho\alpha_{1})
\nonumber \\
&-& \frac{k^{2}e^{k\bar{t}}}{\rho^3}(\delta_{2}\cos\,T + \delta_{1}\sin\,T) +
\frac{k\,\Omega\alpha_1}{\rho^4} + \nonumber \\
\label{eq58}
&+& \frac{e^{k\bar{t}}}{\rho^3}\left(\bar{E}_{2}(\bar{x},\bar{y})\cos\,T
+ \bar{E}_{1}(\bar{x},\bar{y})\sin\,T\right) +   \\
&+&
\frac{1}{\rho^4}\left(\rho\dot\rho(x - \alpha_1) + \rho^{2}\dot\alpha_1
- \Omega\,y - k\rho\,e^{k\bar{t}}(\delta_{1}\cos\,T -
\delta_{2}\sin\,T)\right)\bar{B}(\bar{x},\bar{y}) \,.  \nonumber
\end{eqnarray}

It still remains to take into consideration Faraday's law, which,
in our case, is equivalent to eq. (\ref{eq34}). After a detailed
calculation using the magnetic field (\ref{eq50}) and the
electric field (\ref{eq57}--\ref{eq58}), we find that
Faraday's law imposes
\begin{equation}
\label{eq59} \bar{E}_{2\bar{x}} - \bar{E}_{1\bar{y}} =
k\left(\bar{x}\bar{B}_{\bar{x}} + \bar{y}\bar{B}_{\bar{y}}\right)
\,.
\end{equation}
For $k = 0$ (the Noether point symmetry subcase), equation
(\ref{eq59}) has the general solution
\begin{equation}
\label{eq60}
\bar{E}_1 = - \frac{\partial}{\partial\bar{x}}\bar{V}(\bar{x},\bar{y})
\quad , \quad
\bar{E}_2 = - \frac{\partial}{\partial\bar{y}}\bar{V}(\bar{x},\bar{y})
\quad ,
\end{equation}
where $\bar{V}(\bar{x},\bar{y})$ is an arbitrary function of the
indicated argument. For $k \neq 0$ equation (\ref{eq59})
is a different constraint to be imposed on the electromagnetic field.

In conclusion, we have obtained a very general class of
electromagnetic fields yielding Lie point symmetries. The magnetic
field is given by eq. (\ref{eq50}) and the electric field by eqs.
(\ref{eq57}--\ref{eq58}), together with condition (\ref{eq59}).
The electromagnetic field involves several arbitrary functions,
namely $\rho(t)$, $\alpha_{1}(t)$, $\alpha_{2}(t)$, $\Omega(t)$,
$\bar{B}(\bar{x},\bar{y})$ and $\bar{E_1}(\bar{x},\bar{y})$ or
$\bar{E_2}(\bar{x},\bar{y})$, where $\bar{x}$ and $\bar{y}$ are
defined by eqs. (\ref{eq40}--\ref{eq41}). For instance, for given
$\bar{B}$ and $\bar{E}_1$ the constraint (\ref{eq59}) defines
$\bar{E}_2$ up to the addition of an arbitrary function of
$\bar{y}$.

To conclude this subsection, let us write the equations of motion
in transformed coordinates,
\begin{eqnarray}
\label{eq61}
\bar{x}'' + 2k\bar{x}' + k^{2}\bar{x} &=& \bar{E}_{1}(\bar{x},\bar{y}) +
(\bar{y}' + k\bar{y})\bar{B}(\bar{x},\bar{y}) \,,\\
\label{eq62}
\bar{y}'' + 2k\bar{y}' + k^{2}\bar{y} &=& \bar{E}_{2}(\bar{x},\bar{y}) -
(\bar{x}' + k\bar{x})\bar{B}(\bar{x},\bar{y}) \,.
\end{eqnarray}
As they stand, these equations are not integrable in the general case.

\subsection{The case $\rho = 0$, $k = 0$ and $\Omega \neq 0$}
In this case we have Noether point symmetry. Hence, we simply
quote the main results from reference \cite{Haas}. The
electromagnetic field is given by
\begin{eqnarray}
\label{eqq72} B &=& \bar{B}(\bar{x}, \bar{y}) \,, \\ E_1 &=&
\ddot\beta_1 - \dot\beta_{2}\bar{B}(\bar{x},\bar{y}) + \nonumber
\\ \label{eqq77} &+& (x - \beta_{1})\bar{E}_{1}(\bar{x},\bar{y}) -
(y - \beta_{2})\bar{E}_{2}(\bar{x},\bar{y}) \,,
\\
E_2 &=& \ddot\beta_2 + \dot{\beta}_{1}\bar{B}(\bar{x},\bar{y}) +
\nonumber \\ \label{eqq78} &+& (x -
\beta_{1})\bar{E}_{2}(\bar{x},\bar{y}) + (y -
\beta_{2})\bar{E}_{1}(\bar{x},\bar{y}) \,,
\end{eqnarray}
where $\bar{B}$, $\bar{E}_1$ and $\bar{E}_2$ are arbitrary
functions of $\bar{x}$, $\bar{y}$ given in equations
(\ref{eqq42}--\ref{eqq43}).

Faraday's law requires
\begin{equation}
\label{eqq80} \bar{x}\bar{E}_{2\bar{x}} + 2\bar{E}_2 = -
\bar{B}_{\bar{y}} \,,
\end{equation}
whose solution is
\begin{equation}
\label{eqq81} \bar{E}_2 =
\frac{1}{\bar{x}^2}\frac{\partial\psi}{\partial\bar{y}} \quad ,
\quad \bar{B} = -
\frac{1}{\bar{x}}\frac{\partial\psi}{\partial\bar{x}} \quad ,
\end{equation}
for arbitrary $\psi = \psi(\bar{x},\bar{y})$.

In conclusion, the electromagnetic field is given by eqs.
(\ref{eqq72}--\ref{eqq78}), with the constraint (\ref{eqq81}).
There remains four arbitrary functions, namely
$E_{1}(\bar{x},\bar{y})$, $\psi(\bar{x},\bar{y})$, $\beta_{1}(t)$
and $\beta_{2}(t)$, with $\bar{x}$, $\bar{y}$ defined in equations
(\ref{eqq42}--\ref{eqq43}). We also observe that in the present
case  $\Omega(t)$  has to be chosen constant in order to
produce physically meaningful electromagnetic field (for details,
see \cite{Haas}). Without loss of generality, we take $\Omega = 1$.

\subsection{The case $\rho = 0$, $k = 0$, $\Omega = 0$ and $a_{2} \neq 0$}

Again we have Noether point symmetry. The eletromagnetic fields, from
ref. \cite{Haas}, are
\begin{eqnarray}
\label{eqq83} B &=& \bar{B}(\bar{x}, \bar{y}) \,,\\ \label{eqqq86}
E_1 &=& \frac{\ddot{a}_{1}y}{a_2} -
\frac{\dot{a}_{2}y}{a_2}\bar{B}(\bar{x},\bar{y}) +
\bar{E}_{1}(\bar{x}, \bar{y}) \,,\\ \label{eqqq87} E_2 &=&
\frac{\ddot{a}_{2}y}{a_2} +
\frac{\dot{a}_{1}y}{a_2}\bar{B}(\bar{x},\bar{y}) +
\bar{E}_{2}(\bar{x}, \bar{y}) \,,
\end{eqnarray}
where $\bar{B}$, $\bar{E}_1$ and $\bar{E}_2$ are arbitrary
functions and $\bar{x}$, $\bar{y}$ are defined in equations
(\ref{eqq45}--\ref{eqq46}).

After solving the differential equations arising from Noether's
symmetry condition, we must verify the constraint imposed
by Faraday's law, which, in this case, implies
\begin{eqnarray}
\label{eqq89} \bar{B} &=& \psi_{\bar{x}} \,,\\ \label{eqq91}
\bar{E}_1 &=& - \bar{V}_{\bar{x}} \,,\\ \label{eqq92} \bar{E}_2
&=&
\frac{\ddot{a}_1}{a_2}\bar{x} -
\frac{\dot{a}_2}{a_2}\psi - \psi_{\bar{y}} +
\frac{a_1}{a_2}\bar{V}_{\bar{x}} \,.
\end{eqnarray}
Here, $\psi = \psi(\bar{x}, \bar{y})$ and $\bar{V} =
\bar{V}(\bar{x}, \bar{y})$ are arbitrary functions.

This completely determines this class of solutions for the
electromagnetic field. $B$ is given by eq. (\ref{eqq83}) and $E_1$
and $E_2$ are given by eqs. (\ref{eqqq86}--\ref{eqqq87}). The
functions $\bar{B}$, $\bar{E}_1$ and $\bar{E}_2$, appearing in the
solution, are given by eqs. (\ref{eqq89}--\ref{eqq92}), in terms
of the arbitrary functions $\psi(\bar{x}, \bar{y})$ and
$\bar{V}(\bar{x}, \bar{y})$ with $\bar{x}$, $\bar{y}$ given by
(\ref{eqq45}--\ref{eqq46}). The arbitrary functions $a_{1}(t)$ and
$a_{2}(t)$ are also present in the electromagnetic field, so that
four arbitrary functions participate in the final solution.

\subsection{The case $\rho = 0$ and $k \neq 0$}

In this case the equation for the magnetic field is
\begin{equation}
\label{eq63} B_{\bar{t}} = - 2\dot\Omega(\bar{y}) \,,
\end{equation}
with solution
\begin{equation}
\label{eq64}
B = - 2\dot\Omega(\bar{y})\bar{t} + \bar{B}(\bar{x},\bar{y}) \,.
\end{equation}
Inserting this magnetic field in the equations for the electric
field, yields
\begin{eqnarray}
\label{eq65}
E_{1\bar{t}} &=& k\,E_1 - \Omega\,E_2 +
(\dot\Omega\,x + \dot{a}_{2})(2\dot\Omega\bar{t} - \bar{B}) - \ddot\Omega\,y
+ \ddot{a}_1 \,,\\
\label{eq66}
E_{2\bar{t}} &=& k\,E_2 + \Omega\,E_1 +
(\dot\Omega\,y - \dot{a}_{1})(2\dot\Omega\bar{t} - \bar{B}) + \ddot\Omega\,x
+ \ddot{a}_2 \,.
\end{eqnarray}
This system may be handled with the more convenient variables
\begin{eqnarray}
\label{eq67}
\Sigma_1 &=& e^{-k\bar{t}}(E_{1}\cos\Omega\bar{t} + E_{2}\sin\Omega\bar{t})
\,,\\
\label{eq68}
\Sigma_2 &=& e^{-k\bar{t}}(- E_{1}\sin\Omega\bar{t} + E_{2}\cos\Omega\bar{t})
\,.
\end{eqnarray}
Using these new variables, we have the transformed equations
\begin{equation}
\label{eq69}
\frac{\partial\Sigma_1}{\partial\bar{t}} =
\frac{\partial\psi_1}{\partial\bar{t}} \quad , \quad
\frac{\partial\Sigma_2}{\partial\bar{t}} =
\frac{\partial\psi_2}{\partial\bar{t}} \quad ,
\end{equation}
where
\begin{eqnarray}
\psi_1 &=& (\dot\Omega^{2}\bar{t}^2 -
\dot\Omega\bar{B}\bar{t})\,\cos\bar{x} -
\ddot\Omega\bar{t}\sin\bar{x} + \nonumber \\ \label{eq70} &+&
(2\dot\Omega\dot\gamma_{2}\bar{t} + \ddot\gamma_1 -
\dot\gamma_{2}\bar{B})\,e^{- k\bar{t}}\cos\Omega\bar{t} + \\ &+& (-
2\dot\Omega\dot\gamma_{1}\bar{t} + \ddot\gamma_2 +
\dot\gamma_{1}\bar{B})\,e^{- k\bar{t}}\sin\Omega\bar{t} \,,
\nonumber \\
&\strut& \nonumber \\
\psi_2 &=& (\dot\Omega^{2}\bar{t}^2 -
\dot\Omega\bar{B}\bar{t})\,\sin\bar{x} +
\ddot\Omega\bar{t}\cos\bar{x} + \nonumber \\ \label{eq71} &+& (-
2\dot\Omega\dot\gamma_{1}\bar{t} + \ddot\gamma_2 +
\dot\gamma_{1}\bar{B})\,e^{- k\bar{t}}\cos\Omega\bar{t} \\ &-&
(2\dot\Omega\dot\gamma_{2}\bar{t} + \ddot\gamma_1 -
\dot\gamma_{2}\bar{B})\,e^{- k\bar{t}}\sin\Omega\bar{t} \,.
\nonumber
\end{eqnarray}
The solutions to (\ref{eq69}) are
\begin{equation}
\label{eq72}
\Sigma_1 = \psi_1 + \bar{E}_{1}(\bar{x},\bar{y}) \quad , \quad
\Sigma_2 = \psi_2 + \bar{E}_{2}(\bar{x},\bar{y}) \,,
\end{equation}
The inverse transformation for (\ref{eq67}--\ref{eq68}) is
\begin{eqnarray}
\label{eq73}
E_1 &=& e^{k\bar{t}}(\Sigma_{1}\cos\Omega\bar{t} -
\Sigma_{2}\sin\Omega\bar{t}) \,,\\
\label{eq74}
E_2 &=& e^{k\bar{t}}(\Sigma_{1}\sin\Omega\bar{t} +
\Sigma_{2}\cos\Omega\bar{t}) \,.
\end{eqnarray}
Back in the original coordinates, the resulting electric
field becomes
\begin{eqnarray}
E_1 &=& \ddot\gamma_1 + 2\dot\Omega\dot\gamma_{2}\bar{t} -
\dot\gamma_{2}\bar{B} + \dot\Omega\bar{t}(\dot\Omega\bar{t} -
\bar{B})(x - \gamma_{1}) \nonumber \\ \label{eq75} &-&
\ddot\Omega\bar{t}(y - \gamma_{2}) +
e^{k\bar{t}}\,(\bar{E}_{1}\cos\Omega\bar{t} -
\bar{E}_{2}\sin\Omega\bar{t}) \,,\\ &\strut& \nonumber \\ E_2 &=&
\ddot\gamma_2 - 2\dot\Omega\dot\gamma_{1}\bar{t} +
\dot\gamma_{1}\bar{B} + \dot\Omega\bar{t}(\dot\Omega\bar{t} -
\bar{B})(y - \gamma_{2}) \nonumber \\ \label{eq76} &+&
\ddot\Omega\bar{t}(x - \gamma_{1}) +
e^{k\bar{t}}\,(\bar{E}_{1}\sin\Omega\bar{t} +
\bar{E}_{2}\cos\Omega\bar{t}) \,.
\end{eqnarray}
Here it is more convinient to use a hybrid notation with transformed time
$\bar{t}$ in order to obtain simpler expressions. We should also
stress the generality of the resulting electromagnetic
field, which possesses six arbitrary functions, namely $\gamma_1$,
$\gamma_2$, $\Omega$, $\bar{B}$, $\bar{E}_{1}$ and $\bar{E}_{2}$.

To finalize, the constraint arising from Faraday's law becomes
\begin{eqnarray}
k\,\frac{\partial\bar{B}}{\partial\bar{y}} &=& -
\ddot\Omega(\bar{y}) +
(k\sin\bar{x} - \Omega(\bar{y})\cos\bar{x})\bar{E}_1 -
(k\cos\bar{x} + \Omega(\bar{y})\sin\bar{x})\bar{E}_2 \nonumber
\\
\label{eq77}
&+& (k\cos\bar{x} -
\Omega(\bar{y})\sin\bar{x})\frac{\partial\bar{E}_1}{\partial\bar{x}} +
(k\sin\bar{x} +
\Omega(\bar{y})\cos\bar{x})\frac{\partial\bar{E}_2}{\partial\bar{x}}
\,.
\end{eqnarray}
This condition must be satisfied by the arbitrary functions
appearing in the solution.  For instance, after specifying $\Omega$,
$\bar{E}_1$ and $\bar{E_2}$, we can consider (\ref{eq77}) as an
equation determining $\bar{B}$ up to the addition of an arbitrary
function of $\bar{x}$.

\section{Conclusion}
We have found all classes of electromagnetic fields for which planar
non--relativistic charged particle motion is compatible with Lie
point symmetries. Our procedure is based on the resolution of the
basic system of linear first--order partial differential equations
(\ref{eq26}), (\ref{eq32}--\ref{eq33}) satisfied by the
electromagnetic field, using canonical group variables. As
shown in section 2, there exist four types of canonical group
variables, yielding four classes of electromagnetic fields
compatible with Lie point symmetry. In comparison with the Noether
point symmetry analysis \cite{Haas}, an additional dilatation
invariance term appears in the generator of Lie point symmetries.
This dilatation invariance is associated with an extra category of
electromagnetic fields compatible with Lie point symmetries. The
electromagnetic fields of subsections 4.2 and 4.3 just fit into the
Noether point symmetry case. The electromagnetic field of section
4.1 can be viewed as a natural extension of the Noether point
symmetry case treated in subsection 4.1 of reference \cite{Haas}.
The class shown in section 4.4 of the present work, however, is
essentially new. Its origin can be traced back to the additional
dilatation invariance which is not possible in the Noether's
theorem framework.

In our treatment, we do not include some symmetries corresponding
to excessively particular classes of electromagnetic fields
homogeneous in space.  In this way, we concentrate on classes of
electromagnetic fields depending on arbitrary functions of certain
similarity variables involving space coordinates. These classes may
be useful, for example, in the search for new exact or approximate
solutions for the Vlasov--Maxwell system in colisionless plasma
physics.  Also, as pointed out in the introduction, symmetry may
help reducing the number of relevant coordinates of the problem and
this may represent a considerable reduction in the cost of its
numerical treatment.

\bigskip\noindent
{\bf{\large Acknowledgement}}\\ This work has been partially
supported by Funda\c{c}\~ao de Amparo a Pesquisa do Estado do Rio
Grande do Sul (FAPERGS).

\end{document}